\RequirePackage{fix-cm}
\documentclass{svjour3}                     

\smartqed  

\usepackage{graphicx}
\usepackage{bm}

\newcommand{\be}{\begin{eqnarray}}
\newcommand{\ee}{\end{eqnarray}}
\newcommand{\ba}{\begin{array}}
\newcommand{\ea}{\end{array}}
\newcommand{\bmat}{\left(\begin{array}}
\newcommand{\emat}{\end{array}\right)}
\newcommand{\no}{\nonumber}
\newcommand{\diff}{\mathrm d}
\newcommand{\e}{\mathrm e}
\newcommand{\Tr}{{\rm Tr}\,}

\begin{document}

\title{Full counting statistics and fluctuation--dissipation relation for periodically driven two-state systems}

\titlerunning{Full counting statistics and fluctuation--dissipation relation for periodic systems}

\author{Kazutaka Takahashi \and
Yuki Hino \and
Keisuke Fujii \and
Hisao Hayakawa
}

\authorrunning{K. Takahashi, Y. Hino, K. Fujii, H. Hayakawa}

\institute{K. Takahashi \at
Institute of Innovative Research, Tokyo Institute of Technology, Kanagawa 226--8503, Japan \\
\email{ktaka@qa.iir.titech.ac.jp}
\and
Y. Hino \at
Yukawa Institute for Theoretical Physics, Kyoto University, Kyoto 606--8502, Japan
\at Present address: NTT DATA Mathematical System Inc., 1F Shinanomachi
Rengakan, 35, Shinanomachi, Shinjuku-ku, Tokyo 160--0016, Japan
\and
K. Fujii \at
Department of Physics, Tokyo Institute of Technology, Tokyo 152--8551, Japan \\
iTHEMS Program, RIKEN, Saitama 351--0198, Japan
\and
H. Hayakawa \at
Yukawa Institute for Theoretical Physics, Kyoto University, Kyoto 606--8502, Japan
}

\date{Received: date / Accepted: date}

\maketitle

\begin{abstract}
We derive the fluctuation theorem for a stochastic 
and periodically driven system 
coupled to two reservoirs with the aid of a master equation.
We write down the cumulant generating functions 
for both the current and entropy production in closed compact forms 
so as to treat the adiabatic and 
nonadiabatic contributions systematically.
We derive the fluctuation theorem by taking into account 
the time reversal symmetry and the property 
that the instantaneous currents flowing into the left and the right 
reservoir are not equal.
It is found that the fluctuation--dissipation relation derived from 
the fluctuation theorem involves
an expansion with respect to the time derivative of the affinity.
\end{abstract}

\section{Introduction}

The fluctuation theorem is known 
as one of the most fundamental relations
in nonequilibrium physics~\cite{ECM,GC,EM}. 
It reduces to the standard linear response theory near equilibrium
and holds even when the system is far from equilibrium. 
Using the formal representation of the fluctuation theorem,
we can derive a series of nontrivial relations
such as the fluctuation--dissipation relation and
Onsager's reciprocal relations.

When the system coupled to several reservoirs 
is driven periodically, 
it can be viewed as a heat engine and properties
such as the efficiency can be considered.
Although the second law of thermodynamics gives a universal upper bound 
on the efficiency of heat engines, the bound is not tight 
for systems operated at finite speed.
Since dissipative effects are inevitable in such systems,
it is interesting to consider the problem
from a perspective of nonequilibrium statistical mechanics~\cite{SST}.

The study of this problem is motivated by 
not only fundamental nonequilibrium physics but also 
experimentally-accessible phenomena.
If the system is driven periodically, it causes nontrivial 
dynamical phenomena such as Thouless pumping~\cite{Thouless,NT} 
in which there exists a finite current even in the absence of net bias.
The current has a geometrical interpretation 
and is described by the Berry phase~\cite{Berry} 
even in classical stochastic systems~\cite{SN07}.
This result finds many applications in engineering.
In quantum systems, a periodic protocol induces  
nontrivial effective interactions 
and is used as a method of controlling the system and of finding
nontrivial states~\cite{Magnus,BCOR}.

In this paper, we treat classical stochastic systems 
driven by periodic modulations.
Some previous studies of Thouless pumping processes 
indicated that the presence of the geometrical current leads to 
the violation of the fluctuation theorem,
which originates from the absence of
the Levitov--Lesovik--Gallavotti--Cohen (LLGC) symmetry
in the cumulant generating function of 
full counting statistics~\cite{GC,LL}.
They discussed that the current distribution function
exhibits non-Gaussian fluctuations~\cite{RHL,GAH,WH17,GG,HH}.
Their studies are based on the adiabatic approximation
and it is not clear 
how the nonadiabatic effect affects the results.
We note that the word ``adiabatic'' used in this paper 
is equivalent to quasistatic.
The range of applicability of the standard proof 
of the fluctuation theorem 
is a subtle problem and we need 
a careful treatment to establish the theorem.

The main aim of the present work is 
to study nonadiabatic effects
on the fluctuation theorem
and relations derived from it.
For this purpose, we only focus on a two-level 
system whose master equation involves transition rates 
depending periodically on time.
We use the method of full counting statistics
to study distributions of the current and entropy
production~\cite{LL,LLL,BN,SU,SH}.
The present study is a natural extension of our previous work on 
nonadiabatic effects of the average current~\cite{TFHH}.
The cumulant generating function treated in the full counting statistics 
reflects the underlying symmetry of the system
and is useful in deriving the fluctuation theorem.

Some previous studies suggested a non-Gaussian fluctuation relation 
of the current distribution 
without consideration of the time reversal operation~\cite{RHL,GAH,WH17,GG,HH}. 
The previous approach is directly related to experimental observations, 
but is not adequate to discuss the correct LLGC symmetry relation.
We also show in the present study that 
it is important to introduce the counting fields 
to all junctions to the thermal reservoirs, 
though the previous treatments introduce one counting field. 
Our new approach can capture the process to 
outgoing (or incoming) currents to (from) both the reservoirs.

The paper is organized as follows.
In Sect.~\ref{Sec:system}, we describe the model, 
and summarize previous results.
In Sect.~\ref{Sec:g}, we introduce the full counting statistics 
and define the cumulant generating function. 
It is analyzed by using the dynamical invariant in Sect.~\ref{Sec:di}
and the obtained compact form of the function 
is studied in detail in Sect.~\ref{Sec:gprop}.
We prove the fluctuation theorem in Sect.~\ref{Sec:ft} and 
derive the fluctuation--dissipation relation in Sect.~\ref{Sec:fdr}.
The last section \ref{Sec:conc} is devoted to conclusions.

\section{System}
\label{Sec:system}

We consider a classical stochastic system which couples
to a left and right reservoir.
The system has two possible states, ``empty'' and ``filled'', 
and the dynamics is described by the two-level classical master equation 
\be
 \frac{\diff}{\diff t}|p(t)\rangle = W(t)|p(t)\rangle.
\ee
$|p(t)\rangle$ is a two-component vector
and $W(t)$ is a transition-rate matrix.
The first (second) component of $|p(t)\rangle$ represents the probability 
that the state is empty (filled).
$W(t)$ is decomposed into two parts $W^{\rm (L)}(t)+W^{\rm (R)}(t)$
and each part is parametrized as
\be
 W^{(\mu)}(t)=\bmat{cc}
 -k_{\rm in}^{(\mu)}(t) & k_{\rm out}^{(\mu)}(t) \\
 k_{\rm in}^{(\mu)}(t) & -k_{\rm out}^{(\mu)}(t)
 \emat,
\ee
where $\mu$ represents L or R.
Here, $k_{\rm in}^{(\mu)}(t)$ is the incoming rate from reservoir $\mu$
and $k_{\rm out}^{(\mu)}(t)$ is the corresponding outgoing rate.
They are both assumed to be positive.
We use the following notations throughout this paper: 
\be
 && k_{\rm in}(t)=k_{\rm in}^{\rm (L)}(t)+k_{\rm in}^{\rm (R)}(t), \\
 && k_{\rm out}(t)=k_{\rm out}^{\rm (L)}(t)+k_{\rm out}^{\rm (R)}(t), \\
 && k^{(\mu)}(t)=k_{\rm in}^{(\mu)}(t)+k_{\rm out}^{(\mu)}(t), \\
 && k(t)=k_{\rm in}(t)+k_{\rm out}(t)=k^{\rm (L)}(t)+k^{\rm (R)}(t).
\ee
The matrix form of $W(t)$ is restricted by the conservation
of probability and the above form is the most general one  
in the present two-level system.
Although we basically assume a periodic system
with period $T_0=2\pi/\omega$,
the following analysis is general and can be applied to
arbitrary dynamical systems.

The two-level master equation is solved exactly in Ref.~\cite{TFHH}.
With the initial condition $|p(0)\rangle=(p_0,1-p_0)^{\rm T}$,
the solution is given by 
\be
 |p(t)\rangle 
 = \bmat{c} p_{\rm out}(t)+\delta(t) \\ 1-p_{\rm out}(t)-\delta(t) \emat
 + \left(p_0-p_{\rm out}(0)\right)\e^{-\int_0^t \diff t'\,k(t')}\bmat{c} 1 \\ -1 \emat, \label{pt}
\ee
where $p_{\rm out}(t)=k_{\rm out}(t)/k(t)$ and 
\be
 \delta(t) = -\int_0^t \diff t'\,\dot{p}_{\rm out}(t')
 \e^{-\int_{t'}^{t} \diff t''\,k(t'')}. \label{delta}
\ee
Here, the dot denotes time derivative.
Since the second term on the right hand side of Eq.~(\ref{pt})
decays as $t$ increases, the first term gives 
the state in the long-time limit.
Note that $\delta(t)$ in principle depends  
on the whole history of the system 
and gives rise to nonadiabatic effects.

The exact form of the probability distribution can be used 
to calculate the average current as
\be
 \langle\hat{J}^{(\mu)}\rangle
 =\lim_{T\to\infty}\frac{1}{T_0}\int_{T}^{T+T_0} \diff t\,J_1^{(\mu)}(t), \label{jav}
\ee
where 
\be
 J_1^{(\mu)}(t) = k_{\rm out}^{(\mu)}(t)p_2(t)-k_{\rm in}^{(\mu)}(t)p_1(t),
\ee
We assume in Eq.~(\ref{jav}) that
the system reaches a periodic state
whose period is equal to the modulation period $T_0$.
$p_1$ ($p_2$) denotes the first (second) component of $|p(t)\rangle$.
Using the solution of the master equation, we can write
\be
 && J_1^{\rm (R)}(t)=J_{1{\rm d}}(t)+\frac{k^{\rm (R)}(t)}{k(t)}\left(\dot{p}_{\rm out}(t)+\dot{\delta}(t)\right), \label{J1R}\\
 && J_1^{\rm (L)}(t)=-J_{1{\rm d}}(t)+\frac{k^{\rm (L)}(t)}{k(t)}\left(\dot{p}_{\rm out}(t)+\dot{\delta}(t)\right), \label{J1L}
\ee
where $J_{1{\rm d}}(t)$ represents the dynamical (``classical'')
part of the current as 
\be
 J_{1{\rm d}}(t) =\frac{k_{\rm in}^{\rm (L)}(t)k_{\rm out}^{\rm (R)}(t)
 -k_{\rm out}^{\rm (L)}(t)k_{\rm in}^{\rm (R)}(t)}{k(t)}. \label{Jd}
\ee
The second terms on the right hand side of Eqs.~(\ref{J1R}) and (\ref{J1L}) 
are represented by surface integrals in a parameter space~\cite{TFHH}.
This means that these parts have geometric meaning.
To find current correlations systematically, 
we need to introduce the generating function for the current distribution.
This is the main aim of this study.

\section{Cumulant generating function}
\label{Sec:g}

To calculate statistical quantities, 
we introduce a set of counting fields 
$\chi(t)=(\chi^{\rm (L)}(t),\chi^{\rm (R)}(t))$, where  
$\chi^{\rm (L)}(t)$ is for the left coupling and 
$\chi^{\rm (R)}(t)$ is for the right.
The transition-rate matrix is modified as 
$W(t)\to W(t,\chi(t))$ where 
the diagonal part is unchanged and the off-diagonal part is changed as
\be
 && W_{12}(t) \to k_{\rm out}^\chi(t) := 
 k_{\rm out}^{\rm (L)}(t)\e^{\chi^{\rm (L)}(t)}+k_{\rm out}^{\rm (R)}(t)\e^{\chi^{\rm (R)}(t)}, \\
 && W_{21}(t) \to k_{\rm in}^\chi(t) := 
 k_{\rm in}^{\rm (L)}(t)\e^{-\chi^{\rm (L)}(t)}+k_{\rm in}^{\rm (R)}(t)\e^{-\chi^{\rm (R)}(t)}.
\ee
The explicit form of the counting fields depends  
on the quantity we wish to calculate; several examples 
will be specified.
Here, we only assume that they are real functions of $t$.
Since all the variables can be taken to be real, 
we do not use the standard choice $\e^{i\chi^{(\mu)}(t)}$.
It should be noted that most of the previous studies
on Thouless pumping processes use only $\chi^{\rm (L)}$ or
$\chi^{\rm (R)}$~\cite{RHL,GAH,WH17,GG,HH}.
However, it is important to introduce two counting fields
for both couplings to discuss symmetry relations
such as the fluctuation theorem.

Let us solve the modified master equation 
\be
 \frac{\diff}{\diff t}|p^\chi(t)\rangle 
 = W(t,\chi(t))|p^\chi(t)\rangle, \label{mmaster}
\ee
to calculate the cumulant generating function, which is a functional 
of $\chi(t)$, as  
\be
 g[\chi]=\lim_{T\to\infty}\frac{1}{T}\ln \langle 1|p^\chi(T)\rangle, \label{cgf}
\ee
where $\langle 1|=(1,1)$.
Since $|p^\chi(t)\rangle\to|p(t)\rangle$
and $\langle 1|p^\chi(t)\rangle\to 1$ for $\chi\to 0$,
we have $g[0]=0$.
The current correlation functions can be calculated by differentiating 
$g[\chi]$ with respect to $\chi(t)$.

The definition of the cumulant generating function in Eq.~(\ref{cgf}) 
is a standard choice for systems with time-independent parameters.
We can calculate the long time average of statistical quantities 
by using $g[\chi]$.
In the present case with periodic modulations, 
we are interested in average statistical quantities 
over one cycle as in Eq.~(\ref{jav}).
We show in the next section that 
Eq.~(\ref{cgf}) can be used to calculate the average quantities
when the periodically modulated system approaches a periodic state.

In this study, we consider the following choices of the counting field.
\begin{enumerate}
\item Average current:
\be
 \chi(t)=\chi=(\chi^{\rm (L)},\chi^{\rm (R)}), \label{chiJ}
\ee
where $\chi^{\rm (L)}$ and $\chi^{\rm (R)}$ are independent of time.
The cumulants of current from the system to the left (right) reservoir 
is obtained by differentiation of $g=g(\chi)$ 
with respect to $\chi^{\rm (L)}$($\chi^{\rm (R)}$).
For example, we have
\be
 \left.\frac{\partial}{\partial\chi^{(\mu)}}g(\chi)\right|_{\chi=0}
 =\lim_{T\to\infty}\frac{1}{T}\int_{0}^{T} \diff t\,J_1^{(\mu)}(t).
\ee
When the system approaches a periodic state, 
the initial nonperiodic behavior can be neglected in 
this long-time average 
and this expression converges to Eq.~(\ref{jav}).

\item Entropy production:
\be
 \chi(t) = \left(\chi\ln\left(\frac{k_{\rm out}^{\rm (L)}(t)}{k_{\rm in}^{\rm (L)}(t)}\right),
 \chi\ln\left(\frac{k_{\rm out}^{\rm (R)}(t)}{k_{\rm in}^{\rm (R)}(t)}\right)\right),
 \label{chis}
\ee
where $\chi$ is independent of time.
The entropy production is obtained by
differentiation of $g$ with respect to $\chi$.
This choice of the counting field is based on the fact 
that the entropy production in the reservoir $\mu$
for a process $j\to i$ is given by
$\sigma^{(\mu)}_{ij}=\ln(W_{ij}^{(\mu)}/W_{ji}^{(\mu)})$.

\item Excess entropy production:
 \be
 \chi(t) = \left(\chi\ln\left(\frac{k_{\rm out}(t)}{k_{\rm in}(t)}\right),
 \chi\ln\left(\frac{k_{\rm out}(t)}{k_{\rm in}(t)}\right)\right). \label{chiex}
\ee
By using the steady-state distribution $p^{\rm (s)}$,
the entropy production can be rewritten as 
\be
 \sigma^{(\mu)}_{ij}
 =\ln\frac{W_{ij}^{(\mu)}p_j^{\rm (s)}}{W_{ji}^{(\mu)}p_i^{\rm (s)}}
 +\ln\frac{p_i^{\rm (s)}}{p_j^{\rm (s)}}. \label{exent}
\ee
The first term on the right hand side of Eq.~(\ref{exent}) corresponds to 
heat required to keep the steady state, and 
the rest, the second term, is interpreted as 
the excess entropy production~\cite{OP,HS}.
In the present two-state case,
the detailed balance condition leads to 
\be
 \sigma_{ij}^{\rm ex}=\ln\frac{p_i^{\rm (s)}}{p_j^{\rm (s)}}=\ln\frac{W_{ij}}{W_{ji}},
\ee
which suggests the choice of the counting field 
as shown in Eq.~(\ref{chiex}).
We note that this definition of the excess entropy production 
is different from that in Ref.~\cite{SH}.

\end{enumerate}

\section{Dynamical invariant}
\label{Sec:di}

Our task is to calculate the cumulant generating function
for a given transition rate matrix with $\chi(t)$.
The introduction of the counting field spoils
the property $\langle 1|W=0$, which makes difficult
to solve Eq.~(\ref{mmaster}) in a straightforward manner~\cite{TFHH}.
The main idea of this paper is to use
the dynamical invariant originally introduced 
for quantum harmonic oscillator systems~\cite{LR}.
It is recognized as the fundamental quantity
in the method of shortcuts to adiabaticity~\cite{CRSCGM,STA}.
The extension of the concept to the present system is straightforward and 
we summarize the relevant results below.

A matrix $F^\chi(t)$ is called a dynamical invariant
when it satisfies the relation 
\be
 \frac{\diff F^\chi(t)}{\diff t} = 
 W(t,\chi(t))F^\chi(t) -F^\chi(t) W(t,\chi(t)).
 \label{Ft}
\ee
This quantity has good properties under the only assumption 
that $F^\chi$ and $W$ are diagonalizable~\cite{CRSCGM,STA}.
First, the instantaneous eigenvalues of $F^\chi(t)$ are independent of time.
Second, the solution of the master equation is 
expressed in the most convenient way 
by using the instantaneous eigenstates of $F^\chi(t)$.
We can generally write 
\be
 |p^\chi(t)\rangle
 = \sum_{n=1}^2 C_n
 \e^{\int_0^t \diff t'\,\langle L_n^\chi(t')|W(t',\chi(t'))|R^\chi_n(t')\rangle-\int_0^t \diff t'\,\langle L^\chi_n(t')|\dot{R}^\chi_n(t')\rangle}
 |R^\chi_n(t)\rangle, \label{ptf}
\ee
where $|R^\chi_n(t)\rangle$ represents a right eigenstate of $F^\chi(t)$
and $\langle L^\chi_n(t)|$ represents the corresponding left eigenstate.
$C_n$ is shown to be time independent, which means that 
the solution is given by the adiabatic state with respect to $F^\chi$.
The exponential factor in the second line of Eq.~(\ref{ptf})
represents the geometric ``phase''.
It makes the state invariant under the transformation
\be
 && |R^\chi_n(t)\rangle\to U_n(t)|R^\chi_n(t)\rangle, \\
 && \langle L^\chi_n(t)|\to \langle L^\chi_n(t)|U_n^{-1}(t),
\ee
where $U_n(t)$ represents an arbitrary real function with $U_n(0)=1$.

Since $\Tr F^\chi(t)$ is independent of $t$ and Eq.~(\ref{Ft}) is unchanged 
under the additive and multiplicative changes $F^\chi(t)\to rF^\chi(t)+c$,
where $r$ and $c$ are arbitrary real constants, 
we can 
take $F^\chi(t)$ to be traceless without loss of generality.
Thus, the eigenvalues are found to be $\pm 1$.
To solve Eq.~(\ref{Ft}), we parametrize $F^\chi(t)$ as 
\be
 F^\chi(t)=\bmat{cc} z^\chi(t) & \displaystyle (1+z^\chi(t))\frac{1}{s^\chi(t)} \\ 
 (1-z^\chi(t))s^\chi(t) & -z^\chi(t) \emat. \label{F}
\ee
Then, $s^\chi(t)$ and $z^\chi(t)$ are obtained by solving
\be
 && \dot{s}^\chi=k^\chi_{\rm out}(s^\chi-s^\chi_+)(s^\chi-s^\chi_-), \label{s} \\
 && \dot{z}^\chi = -\left(k^\chi_{\rm out}s^\chi+\frac{k^\chi_{\rm in}}{s^\chi}\right)
 z^\chi
 +k^\chi_{\rm out}s^\chi-\frac{k^\chi_{\rm in}}{s^\chi}, \label{z}
\ee
where
\be
 s^\chi_\pm = \frac{k_{\rm out}-k_{\rm in}
 \pm\sqrt{(k_{\rm out}-k_{\rm in})^2+4k^\chi_{\rm out}k^\chi_{\rm in}}}
 {2k^\chi_{\rm out}}. \label{spm}
\ee
The parametrization in Eq.~(\ref{F}) is convenient
because Eq.~(\ref{s}) for $s^\chi$
is independent of $z^\chi$ and
Eq.~(\ref{z}) for $z^\chi$ is formally solvable.
Using the solutions of $s^\chi$ and $z^\chi$, we can write 
the left and right eigenstates of $F^\chi(t)$ as 
\be
 && \langle L^\chi_n| = \left\{
 \bmat{cc} 1 & \displaystyle
 -\frac{1+z^\chi}{1-z^\chi}\frac{1}{s^\chi} \emat, \quad
 \frac{1}{2}\bmat{cc} 1+z^\chi &
 \displaystyle (1+z^\chi)\frac{1}{s^\chi} \emat
 \right\}, \label{les}\\
 && |R^\chi_n\rangle = \left\{
 \frac{1}{2}\bmat{c} 1-z^\chi \\ -(1-z^\chi)s^\chi \emat, \quad
 \bmat{c} 1 \\
 \displaystyle \frac{1-z^\chi}{1+z^\chi}s^\chi \emat \right\}. \label{res}
\ee
This set of states satisfies the orthonormal relations
$\langle L^\chi_m|R^\chi_n\rangle=\delta_{m,n}$
and the resolution of unity $\sum_n |R^\chi_n\rangle\langle L^\chi_n|=1$.

\begin{center}
\begin{figure}[t]
\begin{center}
\includegraphics[width=0.8\columnwidth]{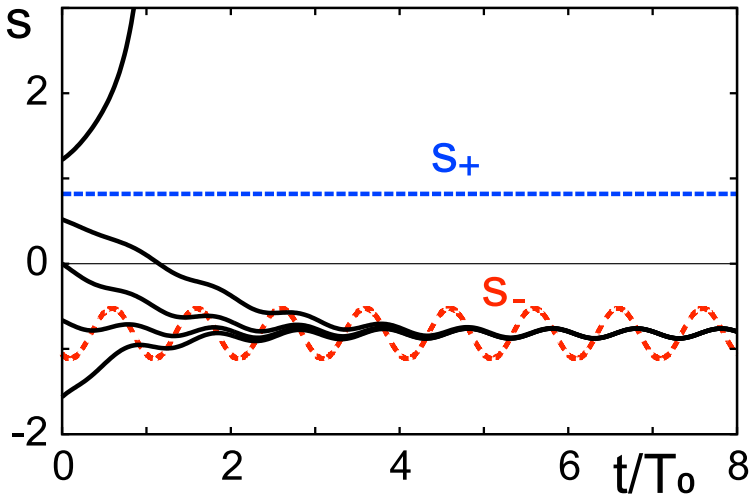}
\end{center}
\caption{Solutions of Eq.~(\ref{s}).
We use the parametrization given in Sect.~\ref{sec-6b} with $\omega=20.0$ 
(in unit of $k_0$ introduced in Eqs.~(\ref{p1})--(\ref{p4}))
and plot the solutions for several choices of the initial condition.
$s^\chi_\pm$ denote fixed points.
For the initial condition $s^\chi(0)< s^\chi_+(0)$, 
$s^\chi(t)$ has asymptotically identical behavior to 
$s^\chi(t)\sim s^\chi_-(t)$.
When $s^\chi$ diverges for $s^\chi(0)> s^\chi_+(0)$,
we can switch to the variable $r^\chi=1/s^\chi$,
which results in the same asymptotic behavior $r^\chi\sim r^\chi_-=1/s^\chi_-$.
}
\label{fig-s}
\end{figure}
\end{center}

Equation (\ref{s}) for $s^\chi$ has two fixed points $s^\chi_{\pm}$.
Since $k^\chi_{\rm out}$ is a nonnegative quantity,  
$s^\chi_+$ is unstable with respect to deviations while $s^\chi_-$ is stable.
Then, we expect that $s^\chi$ approaches the stable fixed point $s^\chi_-$  
as time goes on (See Fig.~\ref{fig-s}).
After transient evolution depending on the initial condition,  
$s^\chi$ relaxes into a state similar to $s^\chi_-$.
The solution of Eq.~(\ref{s}) at large $t$ 
takes negative values and the corresponding solution of $z^\chi$ 
diverges exponentially as we see from Eq.~(\ref{z}).
We note that the fixed point of $z^\chi$ given by 
\be
 z^\chi_0
 =\frac{k^\chi_{\rm out}s_-^\chi-\frac{k^\chi_{\rm in}}{s_-^\chi}}
 {k^\chi_{\rm out}s_-^\chi+\frac{k^\chi_{\rm in}}{s_-^\chi}}
 = -\frac{k_{\rm out}-k_{\rm in}}
 {\sqrt{(k_{\rm out}-k_{\rm in})^2
 +4k^\chi_{\rm out}k^\chi_{\rm in}}},
\ee
describes the adiabatic solution.
The eigenstates of $F^\chi(t)$ in Eqs.~(\ref{les}) and (\ref{res})
become those of $W(t,\chi(t))$
by substituting $s^\chi=s^\chi_-$ and $z^\chi=z^\chi_0$.

To obtain a closed form of the cumulant generating function,
we evaluate $|p^\chi(t)\rangle$ in Eq.~(\ref{ptf}).
We have 
\be
 && \langle L^\chi_n|W|R^\chi_n\rangle
 -\langle L^\chi_n|\dot{R}^\chi_n\rangle
 \no\\
 &=& \left\{
 \ba{l}\displaystyle
 -\frac{1+z^\chi}{1-z^\chi}\frac{k^\chi_{\rm in}}{s^\chi}-k_{\rm in} \\
 \displaystyle
 \frac{1-z^\chi}{1+z^\chi}k^\chi_{\rm out}s^\chi-k_{\rm in}\ea\right. 
 \to \left\{
 \ba{ll}\displaystyle
 \frac{k^\chi_{\rm in}}{s^\chi}-k_{\rm in} & \quad n=1 \\
 -k^\chi_{\rm out}s^\chi-k_{\rm in} & \quad n=2 \ea\right., \label{1steq}
\ee
\be
 \langle 1|R^\chi_n\rangle = \left\{
 \ba{ll}\displaystyle
 \frac{1}{2}(1-z^\chi)(1-s^\chi) & \quad n=1 \\
 \displaystyle
 1+\frac{1-z^\chi}{1+z^\chi}s^\chi & \quad n=2
 \ea 
 \right., \label{2ndeq}
\ee
where the limit $z^\chi\to\infty$ is taken in Eq.~(\ref{1steq}).
Taking the logarithm of Eq.~(\ref{2ndeq}) and
using the solution of $z^\chi$ as 
\be
 z^\chi(T) \sim \exp\left[-\int_0^T\diff t\,
 \left(k^\chi_{\rm out}s^\chi+\frac{k^\chi_{\rm in}}{s^\chi}\right)(t)\right],
\ee
we obtain
\be
 \lim_{T\to\infty}\frac{1}{T}\ln\langle 1|R^\chi_n(T)\rangle = \left\{
 \ba{ll}
 -\lim_{T\to\infty}\frac{1}{T}\int_0^T\diff t\,
 \left(k^\chi_{\rm out}s^\chi+\frac{k^\chi_{\rm in}}{s^\chi}\right)(t) & \quad n=1 \\ 
 0 & \quad n=2
 \ea\right.. \label{2ndeq2}
\ee
Combining Eqs.~(\ref{1steq}) and (\ref{2ndeq2}), we obtain a compact form
\be
 g[\chi]
 &=& \max_n\lim_{T\to\infty}\frac{1}{T}\left[\int_0^T\diff t\,\left(
\langle L^\chi_n|W|R^\chi_n\rangle
 -\langle L^\chi_n|\dot{R}^\chi_n\rangle\right)
 +\ln\langle 1|R^\chi_n(T)\rangle\right] \no\\
 &=& -\lim_{T\to\infty}\frac{1}{T}\int_0^T\diff t\,\left(
 k^\chi_{\rm out}(t)s^\chi(t)+k_{\rm in}(t)\right). \label{g}
\ee
$s^\chi(t)$ is obtained by solving the first-order differential equation
in Eq.~(\ref{s}).
We note that each part of $n=1$ and 2 gives the same expression.

As we see from Fig.~\ref{fig-s}, under periodic modulation, 
the system approaches a periodic state.
Then, we can rewrite Eq.~(\ref{g}) as
\be
 g[\chi]
 = -\lim_{T\to\infty}\frac{1}{T_0}\int_{T}^{T+T_0}\diff t\,\left(
 k^\chi_{\rm out}(t)s^\chi(t)+k_{\rm in}(t)\right), \label{g0}
\ee
which shows that we can calculate statistical quantities
averaged over one cycle by using $g[\chi]$.
The integral is taken over a finite interval and we can find
a geometrical expression as we show in the next section.

\section{Properties of the cumulant generating function}
\label{Sec:gprop}

\subsection{Decomposition}

The cumulant generating function is decomposed into several parts.
We rewrite Eq.~(\ref{s}) for $s^\chi$ as
\be
 s^\chi=s^\chi_--\frac{1}{k^\chi_{\rm out}}\frac{\dot{s}^\chi}{s^\chi_+-s^\chi},
 \label{s2}
\ee
and substitute this into Eq.~(\ref{g}).
Then, $g$ is decomposed as $g=g_{\rm d}+g_{\rm g}$.
The first part is obtained by substituting $s^\chi_-$ into $s^\chi$ as
\be
 g_{\rm d}
 = \frac{1}{T_0}\int_0^{T_0}\diff t\,
 \frac{\sqrt{(k_{\rm out}-k_{\rm in})^2+4k^\chi_{\rm out}k^\chi_{\rm in}}
 -(k_{\rm out}+k_{\rm in})}{2}, \label{gd}
\ee
where we consider a periodic system with the period $T_0$.
This represents the dynamical part and is obtained from
the dynamical ``phase'' $\langle L^\chi_n|W|R^\chi_n\rangle$
in the standard adiabatic treatment.

The other part represents the geometrical part:
\be
 g_{\rm g}
 = \lim_{T\to\infty}\frac{1}{T_0}\int_T^{T+T_0}\diff t\,\frac{\dot{s}^\chi}{s^\chi_+-s^\chi}
 = \lim_{T\to\infty}\frac{1}{T_0}\int_{T}^{T+T_0}\diff t\,\frac{\dot{s}^\chi_+}{s^\chi_+-s^\chi}. \label{gg}
\ee
The adiabatic part is obtained by substituting
the adiabatic solution $s^\chi=s^\chi_-$ into $g_{\rm g}$ as 
\be
 g_{\rm ad} = \frac{1}{T_0}\int_0^{T_0}\diff t\,
 \frac{\dot{s}^\chi_+}{s^\chi_+-s^\chi_-}. 
\ee
This part is obtained from
the geometric ``phase'' $\langle L^\chi_n|\dot{R}^\chi_n\rangle$
in the adiabatic treatment.
The difference between $g_{\rm g}$ and $g_{\rm ad}$ represents
nonadiabatic effects.
The form of $g_{\rm ad}$ leads to a geometric interpretation
as can be seen from the representation 
\be
 g_{\rm ad} = \frac{1}{T_0}\oint\diff \bm{\lambda}\cdot
 \frac{1}{s^\chi_+-s_-^\chi}
 \frac{\partial}{\partial\bm{\lambda}}s^\chi_+,
\ee
where we assume that the time dependence is controlled by
periodic time-dependent parameters $\bm{\lambda}(t)$.
The adiabatic part has a purely geometric interpretation and 
its behavior is characterized by a closed trajectory in the parameter space.
A similar interpretation is possible for the whole geometrical part 
by extending the parameter space to include 
a dynamically generated parameter~\cite{TFHH}.

When the parameters and counting fields are periodic functions 
with period $T_0=2\pi/\omega$, 
$g$ is expanded with respect to $\omega$ as 
\be
 g[\chi]=\sum_{k=0}^\infty \epsilon^k g^{(k)}[\chi], \label{gexp}
\ee
where $\epsilon$ is a dimensionless $\omega$ by a proper scale
($k_0$ in an example of Sect.~\ref{sec-6b})
in $W(t)$.
The dynamical part is given by the zeroth order term $g_{\rm d}=g^{(0)}$,
the adiabatic part is of first order $g_{\rm ad}=\epsilon g^{(1)}$, and 
the rest represents nonadiabatic contributions.
Each part can be found by solving Eq.~(\ref{s2}) iteratively:
\be
 s^\chi
 =s_-^\chi-\frac{\dot{s}^\chi}{k_{\rm out}^\chi(s_+^\chi-s^\chi)}
 =s_-^\chi-\frac{\dot{s}_-^\chi}{k_{\rm out}^\chi(s_+^\chi-s_-^\chi)}+\cdots.
\ee

\subsection{Average current}

The first order term of the cumulant generating function gives the first
moment, i.e., the mean of the statistical quantity.
We derive the average current to confirm that 
the present formulation is consistent with 
previous research~\cite{TFHH}.

We choose $\chi(t)$ as in Eq.~(\ref{chiJ}).
The dynamical part of the current is calculated from Eq.~(\ref{gd}).
We easily find that the expansion to first order in $\chi$
gives $J_{1{\rm d}}(t)$ in Eq.~(\ref{Jd}).
The geometrical part is obtained from 
\be
 \left.\frac{\partial g_{\rm g}(\chi)}{\partial\chi^{(\mu)}}\right|_{\chi=0}
 = \lim_{T\to\infty}\frac{1}{T_0}\int_{T}^{T+T_0} \diff t\,
 \frac{1}{s_+^0-s^0}\frac{\diff}{\diff t}
 \left.\frac{\partial s_+^\chi}{\partial\chi^{(\mu)}}\right|_{\chi=0}.
\ee
At $\chi=0$, $s_+^0=1$ and Eq.~(\ref{s}) can be solved as 
\be
 s^0(t)=1-\frac{1}{p_{\rm out}(t)-p_{\rm out}(0)+\delta(t)+\frac{1}{1-s^0(0)}},
\ee
where $\delta(t)$ is given in Eq.~(\ref{delta}).
We also have
\be
 \left.\frac{\partial s_+^\chi}{\partial\chi^{(\mu)}}\right|_{\chi=0}
 = -\frac{k^{(\mu)}(t)}{k(t)}.
\ee
As a result, we obtain 
\be
 \left.\frac{\partial g_{\rm g}(\chi)}{\partial\chi^{(\mu)}}\right|_{\chi=0}
 = \lim_{T\to\infty}\frac{1}{T_0}\int_{T}^{T+T_0} \diff t\,
 \frac{k^{(\mu)}(t)}{k(t)}
 \left(\dot{p}_{\rm out}(t)+\dot{\delta}(t)\right), 
\ee
where we use the property that $\delta(t)$ 
becomes periodic for long times.
Thus, we conclude that the average current in the present formulation 
gives the known result shown in Eqs.~(\ref{J1R}) and (\ref{J1L}).

\section{Fluctuation theorem}
\label{Sec:ft}

\subsection{Formal considerations}

As mentioned in the introduction,
some studies discussed extended fluctuation relations that
do not satisfy the conventional fluctuation theorem~\cite{RHL,GAH,WH17,GG,HH}.
In the standard analysis of the full counting statistics 
for systems with time-independent parameters,
it is sufficient to introduce the counting field
only for the left, or right, coupling,
because the average currents must satisfy the
relation $\langle\hat{J}^{\rm (L)}\rangle=-\langle\hat{J}^{\rm (R)}\rangle$.
The instantaneous currents, however,  
$J_1^{(\mu)}(t)$ in Eqs.~(\ref{J1R}) and (\ref{J1L})
does not satisfy the simple current conservation, i.e., 
$J_1^{\rm (L)}(t)\ne -J_1^{\rm (R)}(t)$.
To treat the instantaneous currents for both couplings, 
we introduce two counting fields.

It is well known that the fluctuation theorem holds for 
nonequilibrium systems that obey the detailed balance condition.
Therefore, we can expect that the theorem holds in our formulation
by treating the counting field properly.
In this section, we confirm the theorem 
to see how each part of the cumulant generating function 
contributes to the relation.

We investigate the LLGC symmetry of the system~\cite{GC,LL}.
Under the transformation  
\be
 \chi^{(\mu)}(t) \to
 \bar{\chi}^{(\mu)}(t):= -\chi^{(\mu)}(t)-A^{(\mu)}(t), \label{barchi}
\ee
where $A^{(\mu)}(t)$ denotes the affinity 
\be
 A^{(\mu)}(t)= \ln\left(\frac{k_{\rm out}^{(\mu)}(t)}
 {k_{\rm in}^{(\mu)}(t)}\right),
\ee
we find that the transition rate matrix is transformed 
to the transposed matrix as 
\be
 W(t,\chi(t))\to W(t,\bar{\chi}(t))=W^{\rm T}(t,\chi(t)), \label{symW}
\ee
When we use the counting fields
for the entropy production and the excess entropy production
in Eqs.~(\ref{chis}) and (\ref{chiex}),
the transformation is achieved by 
\be
 \chi \to \bar{\chi}=-\chi-1.
\ee

To find the corresponding symmetry of the cumulant generating function,
we write
\be
 g[\chi]=\lim_{T\to\infty}\frac{1}{T}\ln\langle 1|
 \exp_\leftarrow\left(\int_0^T \diff t\,W(t,\chi(t))\right)|p(0)\rangle, \label{g1}
\ee
where $\exp_\leftarrow$ denotes the time-ordered exponential.
This expression can be rewritten as
\be
 g[\chi]=\lim_{T\to\infty}\frac{1}{T}\ln\langle p(0)|
 \exp_\leftarrow
 \left(\int_0^T \diff t\,W^{\rm T}(\bar{t},\chi(\bar{t}))\right)|1\rangle, 
\ee
where $\bar{t}$ is given formally by $\bar{t}=T-t$,
while we can set $\bar{t}=-t$ in periodic systems.
The right vector $|1\rangle$ represents
the simple transpose of the left vector $(\langle 1|)^{\rm T}$ and
similarly for the initial state: $\langle p(0)|=(|p(0)\rangle)^{\rm T}$.
Using Eq.~(\ref{symW}), we obtain
\be
 g[\chi]=\lim_{T\to\infty}\frac{1}{T}\ln\langle p(0)|
 \exp_\leftarrow
 \left(\int_0^T \diff t\,W(\bar{t},\bar{\chi}(\bar{t}))\right)|1\rangle. \label{g2}
\ee
This expression can be analyzed as the original one in Eq.~(\ref{g1}) 
in which the result is insensitive to the choice of the initial condition
as we see from the behavior in Fig.~\ref{fig-s}.
This behavior is consisitent with the observation that 
the system is described by using the largest eigenvalue of 
the transition-rate matrix in the adiabatic limit, 
and of the Floquet effective matrix in the opposite limit~\cite{PBF}.
Comparison of Eq.~(\ref{g2}) with Eq.~(\ref{g1}) gives 
\be
 g[\chi]=g^*[\bar{\chi}], \label{ft}
\ee
where $g^*$ represents the cumulant generating function for
the time-reversed matrix $W(\bar{t},\chi(\bar{t}))$.
To find this symmetric relation, it is crucial to apply the time-reversal 
operation $t\to\bar{t}$ as well as the operation $\chi\to\bar{\chi}$,
which is different from the previous
analysis~\cite{RHL,GAH,WH17,GG,HH}.

Equation (\ref{ft}) is a simple consequence of the symmetry of 
$W(t,\chi(t))$ in Eq.~(\ref{symW}).
The cumulant generating function is transformed into
the current distribution function by the Fourier transformation as
\be
 P[J] = \int [\diff \chi(t)]\,\e^{Tg[i\chi]}
 \exp\left[-i\sum_{\mu={\rm L},\,{\rm R}}
  \int_0^T \diff t\,\chi^{(\mu)}(t) J^{(\mu)}(t)\right], 
\ee
where $J(t)=(J^{\rm (L)}(t),J^{\rm (R)}(t))$ and 
the functional integral is taken over all possible counting fields 
$\chi(t)=(\chi^{\rm (L)}(t),\chi^{\rm (R)}(t))$.
Using Eq.~(\ref{ft}), we find the fluctuation theorem for the current:
\be
 \lim_{T\to\infty}\frac{1}{T}\ln\frac{P[J]}{P^*[-J]}
 =\lim_{T\to\infty}
 \frac{1}{T}\sum_{\mu={\rm L},\,{\rm R}}\int_0^{T} \diff t\,J^{(\mu)}(t)A^{(\mu)}(t).
 \label{ftj}
\ee
Similarly, we can derive the fluctuation theorem for 
the distribution of the entropy production.
Writing the cumulant generating function as $g(\chi)$
where $\chi$ is independent of $t$, 
the distribution function defined as 
\be
 P(\sigma) = \int \frac{\diff \chi}{2\pi}\,\e^{Tg(i\chi)-i\chi\sigma},
\ee
satisfies the known relation
\be
 \ln\frac{P(\sigma)}{P^*(-\sigma)} =\sigma.
\ee

Thus, by careful application of the time-dependent counting fields,
we confirm the existence of standard fluctuation theorem even 
in periodically driven systems such as the Thouless pumping process.
We note that the present discussion 
is applicable to any multi-level system 
provided we can find a suitable transformation $\chi(t)\to\bar{\chi}(t)$.

The present result is different from those in Refs.~\cite{RHL,GAH,WH17,GG,HH}.
The previous works studied a symmetry between $g[\chi^{\rm (L)}]$ and 
$g[\bar{\chi}^{\rm (L)}]$, which leads to a non-Gaussian fluctuation 
relation for $P[J]$.
On the other hand, Eq.~(\ref{ftj}) is the relation between $P[J]$ and $P^*[-J]$.

\subsection{Detailed properties}
\label{sec-6b}

The cumulant generating function $g$ can be expanded
in powers of frequency as in Eq.~(\ref{gexp}),
where each term satisfies Eq.~(\ref{ft}). 
We easily confirm from the explicit form in Eq.~(\ref{gd})
that $g_{\rm d}$ satisfies Eq.~(\ref{ft}).
In Appendix~\ref{app1} we confirm that 
$g_{\rm ad}$ satisfies Eq.~(\ref{ft}).
We also find the relation for each of two operations as 
\be
 g^{(k)*}[\chi]=g^{(k)}[\bar{\chi}]=(-1)^k g^{(k)}[\chi]. \label{gk}
\ee
If $k$ is odd, $g^{(k)}$ introduced in Eq.~(\ref{gexp})
changes the sign under the time-reversal operation.

The above relations hold for any type of counting field.
When we treat the entropy production and the excess entropy production, 
we have additional relations.
In Appendix~\ref{app1}, we show that 
$g_{\rm d}=g_{\rm ad}=0$ for the excess entropy production.
Since the cumulant generating function 
only has a nonadiabatic part, 
the use of the adiabatic approximation does not make sense
when investigating excess entropy production.

\begin{center}
\begin{figure}[t]
\begin{center}
\includegraphics[width=0.8\columnwidth]{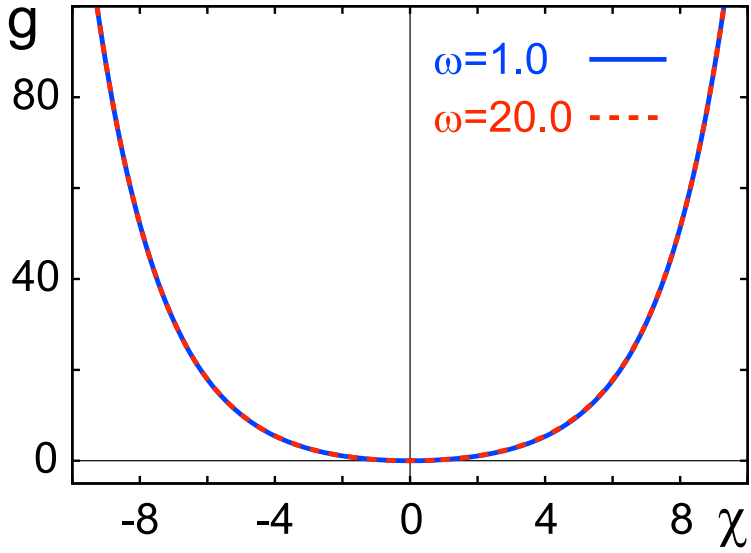}
\end{center}
\caption{
The cumulant generating function for the current.
See Eqs.~(\ref{p1})--(\ref{p4}) on the choice of parameter functions.
All the quantities are plotted in unit of $k_0$.
}
\label{fig-gj}
\end{figure}
\end{center}

\begin{center}
\begin{figure}[t]
\begin{center}
\includegraphics[width=0.8\columnwidth]{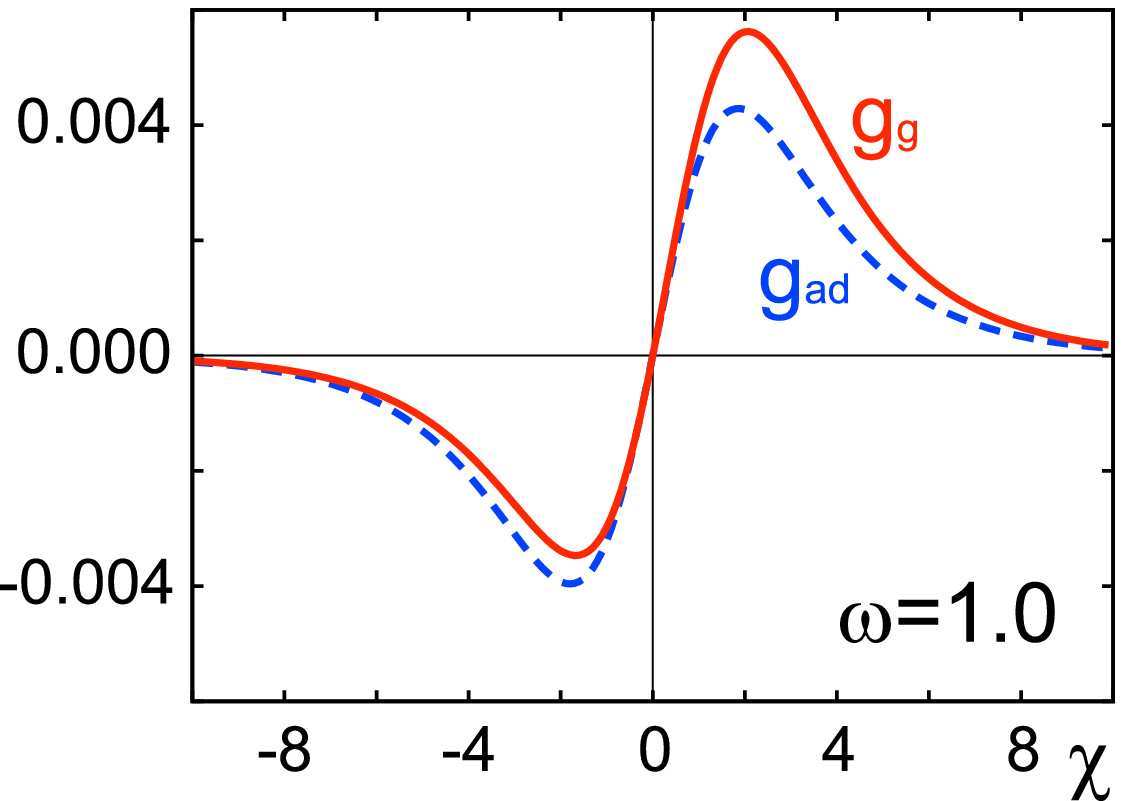}
\includegraphics[width=0.8\columnwidth]{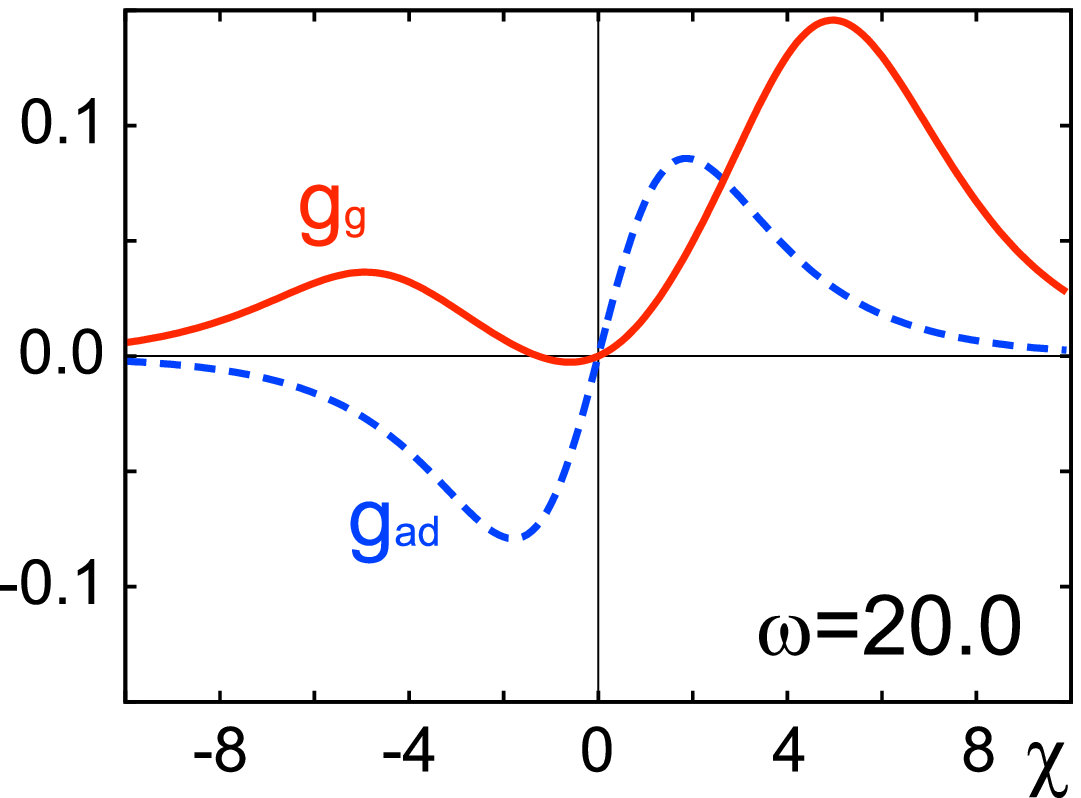}
\end{center}
\caption{
The geometrical part of 
the cumulant generating function 
$g_{\rm g}=\sum_{k=1}^\infty \epsilon^kg^{(k)}$ for the current.
$g_{\rm ad}=\epsilon g^{(1)}$ is the first term of the expansion
and represents the adiabatic contribution.
}
\label{fig-gjg}
\end{figure}
\end{center}

\begin{center}
\begin{figure}[t]
\begin{center}
\includegraphics[width=0.8\columnwidth]{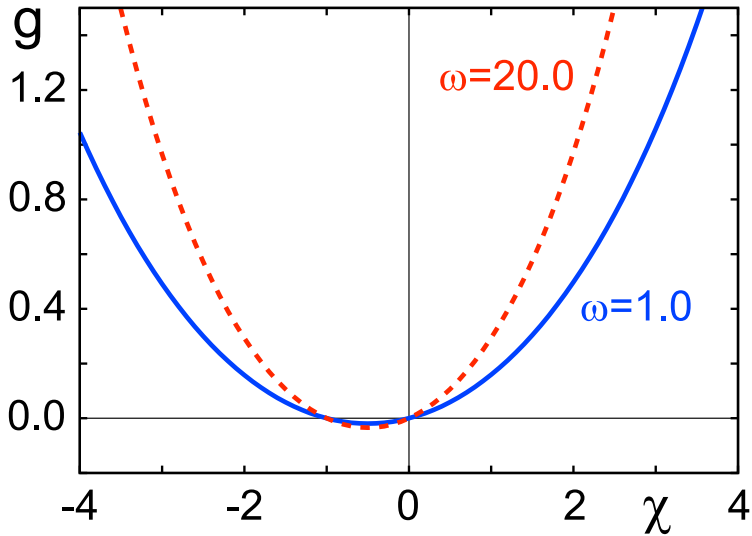}
\end{center}
\caption{
The cumulant generating function for the entropy production.
}
\label{fig-gs}
\end{figure}
\end{center}

\begin{center}
\begin{figure}[t]
\begin{center}
\includegraphics[width=0.8\columnwidth]{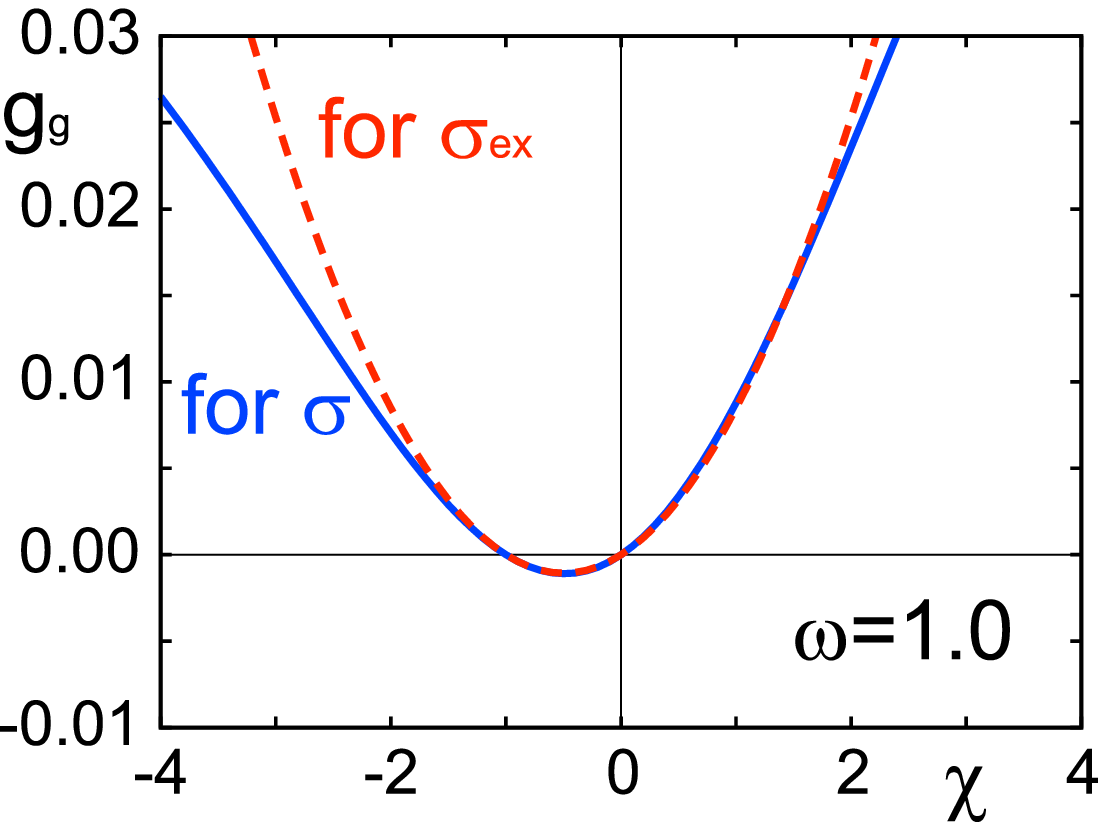}
\includegraphics[width=0.8\columnwidth]{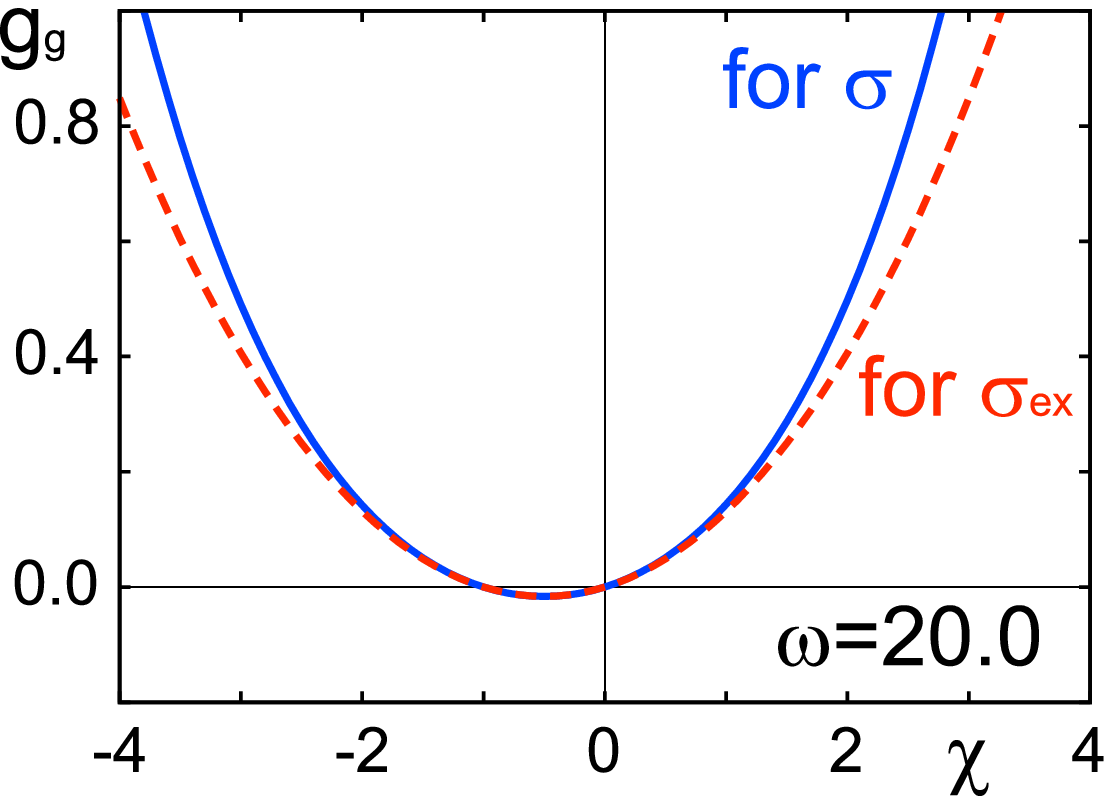}
\end{center}
\caption{
The geometrical (nonadiabatic) part of 
the cumulant generating function for the entropy production (solid lines)
and the excess entropy production (dashed lines).
}
\label{fig-gsg}
\end{figure}
\end{center}

We plot the cumulant generating functions 
in Figs.~\ref{fig-gj}--\ref{fig-gsg} for various cases.
We follow Ref.~\cite{SN07} on the choice of parametrizations 
and take 
\be
 && k_{\rm in}^{\rm (L)}(t)=k_0\left(1+r_1\cos\omega t\right), \label{p1}\\
 && k_{\rm in}^{\rm (R)}(t)=k_0\left(1+r_2\sin\omega t\right), \label{p2}\\
 && k_{\rm out}^{\rm (L)}(t)=k_0, \label{p3}\\
 && k_{\rm out}^{\rm (R)}(t)=k_0, \label{p4}
\ee
with $r_1=0.6$ and $r_2=0.4$.
We take $(\chi^{\rm L}(t),\chi^{\rm R}(t))=(\chi,0)$
for the counting field of the current.

In the case of the current, the dynamical part 
which is independent of $\omega$ is a dominant contribution
and the geometrical part takes rather small values 
as we see in Figs.~\ref{fig-gj} and \ref{fig-gjg}.
Therefore, weak $\omega$ dependence of the cumulant generating function
in Fig.~\ref{fig-gj} is dominated by the dynamical part.
The dynamical part is canceled out for the average current over one cycle
and the geometric part plays the dominant role in that case.
When the frequency is small,  
the difference between $g_{\rm g}$ and $g_{\rm ad}$ is small,
which means that the adiabatic approximation gives accurate results.
The difference becomes large for large frequencies.
We note that the slope at $\chi=0$ represents the average current 
$\langle\hat{J}^{\rm (L)}\rangle$.
The current is significantly suppressed for large values of $\omega$.
This result is consistent with that in Ref.~\cite{TFHH}.
We also confirm that Eq.~(\ref{ft}) holds.

The cumulant generating functions for the entropy production
and the excess entropy production 
are plotted in Figs.~\ref{fig-gs} and \ref{fig-gsg}.
We find that $g(0)=g(-1)=0$ should be satisfied.
We also see that $g(\chi)$ grows rapidly as the frequency increases
in both cases.

\section{Fluctuation--dissipation relations}
\label{Sec:fdr}

\subsection{General formula}

Having confirmed that the fluctuation theorem, Eq.~(\ref{ft}), 
holds in our formulation, let us clarify its physical implications.
The fluctuation theorem gives an infinite series of nontrivial relations such 
as the fluctuation--dissipation relations and 
Onsager's reciprocal relations, which can be examined experimentally.
Our main interest is to find formulae for periodically-modulated systems 
involving nonadiabatic effects.

Generally, the cumulant generating function for the current is expanded as 
\be
 && g[\chi] =
 \frac{1}{T}\int_0^T\diff t\,\chi^{(\mu)}(t)J_1^{(\mu)}(t)
 +\frac{1}{2T}\int_0^T\diff t_1\diff t_2\,\chi^{(\mu)}(t_1)J_2^{(\mu\nu)}(t_1,t_2)\chi^{(\nu)}(t_2) \no\\ &&
 +\frac{1}{6T}\int_0^T\diff t_1\diff t_2\diff t_3\,\chi^{(\mu)}(t_1)\chi^{(\nu)}(t_2)\chi^{(\lambda)}(t_3)J_3^{(\mu\nu\lambda)}(t_1,t_2,t_3) +\cdots, \label{gchi}
\ee
where the summation is taken over repeated indices.
$\{J_k^{(\mu_1,\dots,\mu_k)}(t_1,\dots,t_k)\}_{k=1,2,\dots}$ 
denotes a set of local current correlators to be obtained from
the cumulant generating function 
which satisfies the fluctuation theorem in Eq.~(\ref{ft}).
For example, by comparing the order of $\chi$ 
for the relation $g[\bar{\chi}]=g^*[\chi]$, 
where $\bar{\chi}$ is defined in Eq.~(\ref{barchi}), 
we obtain 
\be
 && -\frac{1}{T}\int_0^T\diff t\,A^{(\mu)}(t)J_1^{(\mu)}(t)
 \no\\ &&
 +\frac{1}{2T}\int_0^T\diff t_1\diff t_2\,A^{(\mu)}(t_1)J_2^{(\mu\nu)}(t_1,t_2)A^{(\nu)}(t_2)+\cdots=0
\ee
at the zeroth order, and 
\be
 && J_1^{(\mu)}(t)+(J_1^{(\mu)}(t))^*
 = \int_0^T\diff t'\,J_2^{(\mu\nu)}(t,t')A^{(\nu)}(t') \no\\ &&
 -\frac{1}{2}\int_0^T\diff t'\diff t''\,
 J_3^{(\mu\nu\lambda)}(t,t',t'')A^{(\nu)}(t')A^{(\lambda)}(t'') +\cdots, \label{fdr0}
\ee
at the first order.
$(J_1^{(\mu)}(t))^*$ represents the local average current 
of the time-reversed protocol.
Equation (\ref{fdr0}) corresponds to the fluctuation-dissipation relation
in our system, which 
we study the detailed structure of the relation in the following.

Since the cumulant generating function is expanded 
in frequency $\omega$ as in Eq.~(\ref{gexp}), 
Eq.~(\ref{ft}) holds for each of the dynamical, adiabatic, 
and nonadiabatic parts.
Correspondingly $J_k^{(\mu_1,\dots,\mu_k)}(t_1,\dots,t_k)$ is decomposed 
into three parts as $J_{k}=(J_k)_{\rm d}+(J_k)_{\rm ad}+(J_k)_{\rm nad}$.
We can study each part separately.

\subsection{Dynamical part}

The dynamical part of the cumulant generating function is obtained 
in Eq.~(\ref{gd}).
It is represented by an integral
over a single variable $t$ and is a functional of 
$\chi^{\rm (R)}(t)-\chi^{\rm (L)}(t)$ as 
\be
 g_{\rm d}[\chi] =
 \frac{1}{T_0}\int_0^{T_0}\diff t\,
 \sum_{k=1}^\infty \frac{1}{k!}J_{k{\rm d}}(t)\left(\chi^{\rm (R)}(t)-\chi^{\rm (L)}(t)\right)^k. \label{gchid}
\ee
where $J_{k{\rm d}}(t)$ represents the local current correlator at order $k$.
The explicit form of $J_{1{\rm d}}(t)$ is given in Eq.~(\ref{Jd}).
This cumulant generating function satisfies the relation 
$g_{\rm d}[\bar{\chi}]=g_{\rm d}[\chi]=g_{\rm d}^*[\chi]$.
We note that the function is invariant under the time-reversed protocol.
Then, by comparing coefficients of $g_{\rm d}[\bar{\chi}]$ and $g_{\rm d}[\chi]$ 
at $k$th order in the counting field, we obtain 
\be
 \left(1-(-1)^k\right)J_{k{\rm d}}(t)=\sum_{\ell=1}^\infty
 \frac{(-1)^{\ell+1}}{\ell!}J_{k+\ell,{\rm d}}(t)(A^{(-)}(t))^{\ell}, \label{fdr-d}
\ee
where $A^{(-)}=A^{\rm (R)}-A^{\rm (L)}$.
It gives the standard fluctuation--dissipation relation 
by setting $k=1$~\cite{HH}.
We can write down a similar relation for the current correlation 
at odd order.
We can also write down the derivative form 
\be
 2\left.\frac{\delta}{\delta A^{(-)}(t)}\frac{1}{T_0}\int_0^{T_0} \diff t'\,J_{2k-1,{\rm d}}(t')\right|_{A^{(-)}=0}=\left.\frac{1}{T_0}J_{2k,{\rm d}}(t)\right|_{A^{(-)}=0}, \label{fdr-d2}
\ee
where $k=1,2,\dots$. 

In the dynamical part, 
the correlations are local in $t$ and 
the relations are written by using $A^{(-)}$.
Since the dynamical part is obtained by neglecting nonadiabatic effects,
these results coincide with those in static systems.

\subsection{Geometrical part}

Let us study the fluctuation-dissipation relations for the geometrical part
to see  how Eq.~(\ref{fdr-d}) is changed.

The cumulant generating function is expanded in frequency and 
each order satisfies the relation in Eq.~(\ref{gk}).
For example, the adiabatic current changes its sign by considering
the time-reversed protocol as $(J^{(\mu)}(t))^*=-J^{(\mu)}(t)$.
Thus, in the case of the adiabatic part, 
it is impossible to relate the average current 
to higher-order correlations as in Eq.~(\ref{fdr0}).
This is contrasted to the analysis of Ref.~\cite{HH} where 
the fluctuation-dissipation relation was discussed on the adiabatic part.
To find nontrivial corrections to Eq.~(\ref{fdr0}), 
we need to consider higher order correlations or 
to go beyond the adiabatic approximation.

To obtain a nontrivial contribution 
which cannot be found in the dynamical part, 
we examine the adiabatic part.
The adiabatic part of the cumulant generating function is given  
in Eq.~(\ref{gad-exp}) in Appendix~\ref{app1}.
It is rewritten as 
\be
 && g_{\rm ad}[\chi]= \frac{1}{2T_0}\int_0^{T_0}\diff t\,\frac{1}
 {\sqrt{(k_{\rm out}-k_{\rm in})^2+4k_{\rm out}^\chi k_{\rm in}^\chi}} \no\\
 &&\times\Biggl[
 \dot{k}_{\rm out}-\dot{k}_{\rm in}
 -\frac{\dot{k}_{\rm out}^{\rm (L)}+\dot{k}_{\rm out}^{\rm (R)}\e^{-\chi^{(-)}}
 +\frac{1}{2}\dot{\chi}^{(-)}
 (k_{\rm out}^{\rm (L)}-k_{\rm out}^{\rm (R)}\e^{-\chi^{(-)}})}
 {k_{\rm out}^{\rm (L)}+k_{\rm out}^{\rm (R)}\e^{-\chi^{(-)}}}
 (k_{\rm out}-k_{\rm in})
 \no\\ && 
 -\frac{1}{2}\dot{\chi}^{(+)}(k_{\rm out}-k_{\rm in})
 \Biggr] 
\ee
where $\chi^{(\pm)}=\chi^{\rm (R)}\pm\chi^{\rm (L)}$.
Since $k_{\rm out}^\chi k_{\rm in}^\chi$ is a function of $\chi^{(-)}$, 
this expression is expanded as 
\be
 && g_{\rm ad}[\chi]=\frac{1}{T_0}\int_0^{T_0}\diff t\,
 \sum_{k=1}^\infty \frac{1}{k!}\left[
 \dot{\bm{\lambda}}
 \cdot\bm{a}_{k}(\bm{\lambda}(t))
 \left(\chi^{(-)}(t)\right)^k 
 \right.\no\\ && \left.
 -kq_{k}(\bm{\lambda}(t))\dot{\chi}^{(+)}(t)
 \left(\chi^{(-)}(t)\right)^{k-1}\right].
\ee
to define the coefficients $\{\bm{a}_k\}_{k=1,2,\dots}$
and $\{q_k\}_{k=1,2,\dots}$.
Here, $\bm{\lambda}(t)=(\lambda_1(t),\lambda_2(t),\dots)$ represents a set of
time-dependent controllable parameters in the model.
They are periodic functions with the period $T_0$
as well as the counting fields $\chi^{(\pm)}(t)$.
The adiabatic part involves the first-order derivative.
The $k$th order local current correlation is given by 
\be
 && J_{k,{\rm ad}}^{\rm (R)}(t)=\dot{\bm{\lambda}}(t)\cdot\left(
 \bm{a}_k(\bm{\lambda}(t))+\frac{\partial}{\partial\bm{\lambda}}q_k(\bm{\lambda}(t))\right), \\
 && J_{k,{\rm ad}}^{\rm (L)}(t)=(-1)^k\dot{\bm{\lambda}}(t)\cdot\left(
 \bm{a}_k(\bm{\lambda}(t))-\frac{\partial}{\partial\bm{\lambda}}q_k(\bm{\lambda}(t))\right).
\ee
Taking the average over one cycle, we obtain 
\be
 J_{k,{\rm ad}}^{\rm (R)}=
 (-1)^kJ_{k,{\rm ad}}^{\rm (L)}=
 \frac{1}{T_0}\int_0^{T_0} \diff t\,\dot{\bm{\lambda}}(t)\cdot
 \bm{a}_k(\bm{\lambda}(t)).
\ee
$\dot{q}_{k}(\bm{\lambda})$ contributes only to the local current and 
can be observed by measuring 
the difference between the right and left currents.

By using the relation $g_{\rm ad}[\bar{\chi}]=-g_{\rm ad}[\chi]$, 
we obtain at $k$th order
\be
 && \left(1+(-1)^k\right)\dot{\bm{\lambda}}\cdot\bm{a}_k
 \no\\ &&
 = \sum_{\ell=1}^\infty\frac{(-1)^{\ell+1}}{\ell!}\left(
 \dot{\bm{\lambda}}\cdot\bm{a}_{k+\ell} A^{(-)}
 -\ell q_{k+\ell} \dot{A}^{(+)}
 \right)
 (A^{(-)})^{\ell-1}, \label{fdr-ada}\\
 && \left(1+(-1)^k\right)q_k=\sum_{\ell=1}^\infty\frac{(-1)^{\ell+1}}{\ell!}
 q_{k+\ell} (A^{(-)})^{\ell}. \label{fdr-adq}
\ee
These relations are basically 
written by using $A^{(-)}$ as in the dynamical part.
The difference is that the time derivative of $A^{(+)}=A^{\rm (R)}+A^{\rm (L)}$ 
appears in the expansion.
The property that the time derivative of the parameters appears
in the expansion is expected from the general consideration
of the adiabatic response~\cite{LBOA,BS,HH2}.
Our formulation is self-contained and naturally leads to those results.

Higher-order parts of the cumulant generating function, $g^{(k)}[\chi]$
with $k\ge 2$, can be analyzed in a similar way.
The even-order parts give corrections to Eq.~(\ref{fdr-d})
and the odd-order to Eqs.~(\ref{fdr-ada}) and (\ref{fdr-adq}).
The result is expanded with respect to $\dot{A}^{(+)}$ 
as well as $A^{(-)}$ and the explicit form can be extracted from Eq.~(\ref{gg}).

\section{Conclusion}
\label{Sec:conc}

In conclusion, we obtained the cumulant generating
function for two-level stochastic systems
and derived a series of fluctuation relations applicable to
periodically driven systems converging to a periodic state.
Our findings are summarized as follows.

\begin{itemize}

\item
All of our results are derived from a compact form of
the cumulant generating function shown in Eq.~(\ref{g}).
We derive various useful expressions 
by using the properties of the dynamical invariant.
Although we need to solve the differential equation in Eq.~(\ref{s}),
the equation is simple and a systematic treatment is possible.
In fact, we can extract the adiabatic part and the nonadiabatic part
from the expression and can study how each part contributes to the result.

\item 
The cumulant generating function is useful not only
for calculating the current distribution but also
for finding the underlying symmetry.
To derive the fluctuation theorem, Eq.~(\ref{ft}), 
we stress that introducing the instantaneous counting field 
for coupling to each reservoir is important.
It allows us to consider the symmetry under
the transformation in Eq.~(\ref{barchi}).

\item
We also stress that the time-reversal operation is crucial 
in deriving the fluctuation theorem.
Previous studies on the non-Gaussian fluctuation relation 
paid attention to a relation between $P[J]$ and $P[-J]$,
rather than $P[J]$ and $P^*[-J]$.
We showed that, even within the adiabatic approximation,
the effect of the time-reversal operation is important.

\item 
Equation (\ref{ft}) gives a series of nontrivial relations
such as Eqs.~(\ref{fdr-d}), (\ref{fdr-ada}), and (\ref{fdr-adq}).
The result is expanded with respect
to $\dot{A}^{(+)}(t)$ as well as $A^{(-)}(t)$.

\end{itemize}

The most important result of this paper is a systematic method of 
dealing with the nonadiabatic effects with desired accuracy.
It is an interesting problem to generalize the present method 
to multistate systems.
The generalization for the current was discussed in Ref.~\cite{TFHH}.
In principle, the decomposition of the function into 
the dynamical and geometrical parts 
is possible even for general cases.
It is generally a difficult task to handle the equation
for the dynamical invariant and we must use a different method 
to find a compact form of the cumulant generating function.
This problem will be discussed in a forthcoming paper.

\begin{acknowledgements}
We are grateful to Ken Funo for useful discussions and comments.
We also thank Ville Paasonen for his critical reading of the manuscript.
This work was supported by JSPS KAKENHI Grant 
Number JP19J13698 (K.~F.) and Number JP16H04025 (H.~H. and Y.~H.).
The part of this study is supported by Ishizue 2020 by 
Kyoto University Research Development Program.
K.~T. acknowledges the warm hospitality of the Yukawa Institute for 
Theoretical Physics, Kyoto University during his stay there 
to promote the collaboration among the authors.
\end{acknowledgements}

\appendix
\section{On the adiabatic part of the cumulant generating function}
\label{app1}

\subsection{General properties}

We consider the adiabatic part of the cumulant generating function 
\be
 g_{\rm ad}[\chi]
 = \frac{1}{T_0}\int_0^{T_0}\diff t\,
 \frac{\dot{s}^\chi_+}{s^\chi_+-s^\chi_-}
 = \frac{1}{2T_0}\int_0^{T_0}\diff t\,
 \frac{\dot{s}^\chi_++\dot{s}^\chi_-}{s^\chi_+-s^\chi_-}. \label{A1}
\ee
Using Eq.~(\ref{spm}), we rewrite Eq.~(\ref{A1}) as
\be
 g_{\rm ad}[\chi] = \frac{1}{2T_0}\int_0^{T_0}\diff t\,\left[
 \frac{\dot{k}_{\rm out}-\dot{k}_{\rm in}}{\sqrt{(k_{\rm out}-k_{\rm in})^2+4k_{\rm out}^\chi k_{\rm in}^\chi}}
 -\frac{\dot{k}_{\rm out}^\chi}{k_{\rm out}^\chi}
 \frac{k_{\rm out}-k_{\rm in}}{\sqrt{(k_{\rm out}-k_{\rm in})^2+4k_{\rm out}^\chi k_{\rm in}^\chi}}
 \right]. \no\\ \label{gad-exp}
\ee
To show that this function satisfies the relation in Eq.~(\ref{ft}), 
we apply the time-reversal operation $g\to g^*$ 
and the transformation of the counting field $\chi\to\bar{\chi}$.
We have 
\be
 g^*_{\rm ad}[\bar{\chi}] = -\frac{1}{2T_0}\int_0^{T_0}\diff t\,\left[
 \frac{\dot{k}_{\rm out}-\dot{k}_{\rm in}}{\sqrt{(k_{\rm out}-k_{\rm in})^2+4k_{\rm out}^\chi k_{\rm in}^\chi}}
 -\frac{\dot{k}_{\rm in}^\chi}{k_{\rm in}^\chi}
 \frac{k_{\rm out}-k_{\rm in}}{\sqrt{(k_{\rm out}-k_{\rm in})^2+4k_{\rm out}^\chi k_{\rm in}^\chi}}
 \right]. \no\\
\ee
We note that the combination $k_{\rm out}^\chi k_{\rm in}^\chi$ is 
invariant under the replacement $\chi\to\bar{\chi}$.
Then, taking the difference between $g_{\rm ad}[\chi]$ and 
$g_{\rm ad}^*[\bar{\chi}]$, we obtain
\be
 &&
 g_{\rm ad}[\chi]-g^*_{\rm ad}[\bar{\chi}]
 \no\\
 &=& \frac{1}{T_0}\int_0^{T_0}\diff t\,\left[
 \frac{\dot{k}_{\rm out}-\dot{k}_{\rm in}}{\sqrt{(k_{\rm out}-k_{\rm in})^2+4k_{\rm out}^\chi k_{\rm in}^\chi}}
 -
 \frac{k_{\rm out}-k_{\rm in}}{2\sqrt{(k_{\rm out}-k_{\rm in})^2+4k_{\rm out}^\chi k_{\rm in}^\chi}}\frac{\diff}{\diff t}\ln (k_{\rm out}^\chi k_{\rm in}^\chi)
 \right]. \no\\
\ee
We change the variables according to 
\be
 && k_{\rm out}-k_{\rm in}=r\cos\theta, \\
 && \sqrt{4k_{\rm out}^\chi k_{\rm in}^\chi}=r\sin\theta,
\ee
to find
\be
 g_{\rm ad}[\chi]-g^*_{\rm ad}[\bar{\chi}]
 = \frac{1}{T_0}\int_0^{T_0}\diff t\,\frac{\dot{\theta}}{\sin\theta} = 0.
 \label{gad0}
\ee
We note that $0<\theta<\pi$ in the present parametrization 
with $k_{\rm out}^\chi k_{\rm in}^\chi>0$ and 
the integral is not divergent.

\subsection{Cumulant generating function for the excess entropy production}

When we consider the excess entropy production, 
we choose the counting field as shown in Eq.~(\ref{chiex}).
The cumulant generating function in that case is written as 
\be
 g_{\rm ad}(\chi) 
 = \frac{1}{2T_0}\int_0^{T_0} \diff t\,
 \frac{\left(\dot{k}_{\rm out}k_{\rm in}-\dot{k}_{\rm in}k_{\rm out}\right)
 \left[k_{\rm in}-\chi(k_{\rm out}-k_{\rm in})\right]}{kk_{\rm out}k_{\rm in}}. 
\ee
Then, by changing variables according to 
$k_{\rm out}=r\cos\theta$, $k_{\rm in}=r\sin\theta$,
we find that the integrand depends only on $\theta$
as in Eq.~(\ref{gad0}).
Then, the adiabatic part of the cumulant generating function is
identically zero.


\end{document}